\documentclass[12pt]{article} \textheight 23cm
\usepackage{graphicx}
\usepackage{amsmath}
\usepackage{color}
\begin{document} \begin{center}
{\large Crack Formation in Laponite Gel under AC Fields     }\\ \vskip 0.5cm
Tajkera Khatun$^1$, Tapati Dutta$^2$ and Sujata Tarafdar$^1$\\
\vskip 0.5cm

$^1$ Condensed Matter Physics Research Centre, Physics Department, Jadavpur University, Kolkata 700032, India\\

$^2$ Physics Department, St. Xavier's College, Kolkata 700016, India\\

\end{center}
\vskip .5cm
\noindent {\bf Abstract}\\ \vskip .5cm
Crack formation patterns in laponite gel are known to be strongly affected by DC electric fields. We show that AC fields produce equally remarkable patterns in a radially symmetric set-up. The character of the pattern depends crucially on the field strength. A significant feature observed is the bending of radial cracks, with the curvature increasing as field strength is increased. Fields of 20 to 70 V have been applied and several features of the resulting patterns quantified. Striations on the fracture surfaces and crack speeds are also studied.
\vskip .5cm

The importance of the study of desiccation cracks and fracture growth as an interdisciplinary field is widely accepted, as evident from the considerable literature being published in journals of physics, geology, mechanical and chemical engineering and mathematics. More recently it has been realized that crack patterns can be controlled and tailored to produce `designer cracks' which may be useful in nano-fabrication\cite{nature} and manufacture of surfaces with special texture. Exposure to electric fields may be one of the methods for controlling fracture.
Formation of desiccation crack patterns in the synthetic clay - laponite, is strongly affected by the presence of an electric field\cite{mal,tajkera}. We show here that AC fields produce equally remarkable effects. Clay particles are charged in aqueous environment\cite{book}, so response to an electric field is quite natural. The effect of electric fields on mechanical fracture in dielectric materials was reported much earlier\cite{suo}, in particular an AC field seemed to favour crack growth normal to the field direction\cite{cao-ac}.

In the present communication we report preliminary results on the effect of radially symmetric AC fields of varying strength, on a laponite gel dried in a circular polypropylene petri-dish of 10 cm diameter. As the field strength is increased from 20 V to 70 V AC (50 Hz), the pattern of cracks undergoes a remarkable series of changes which are quite reproducible.
The set up is similar to the earlier studies\cite{tajkera}, aluminium foils are used to make the electrodes, one in the form of a thin rod, is placed at the center and the counter-electrode is fitted along the inner periphery of the petri-dish in the form of a short cylinder.
In the present experiments 2.5 gm of laponite(Rockwood additives) is added to 40 ml of distilled water containing a trace of crystal violet dye(to make the cracks easily visible), while stirring on a magnetic stirrer. The mixture is stirred for 20-30 sec and poured in  circular petri dishes of 10 cm diameter fitted with two electrodes. After waiting for 5 minutes till the solution spreads out evenly and gels, a variac from the main supply is used to apply an AC field between the two electrodes while the gel is allowed to dry. 

\begin{figure}[h]
\begin{center}
\includegraphics[width=12.0cm, angle=0]{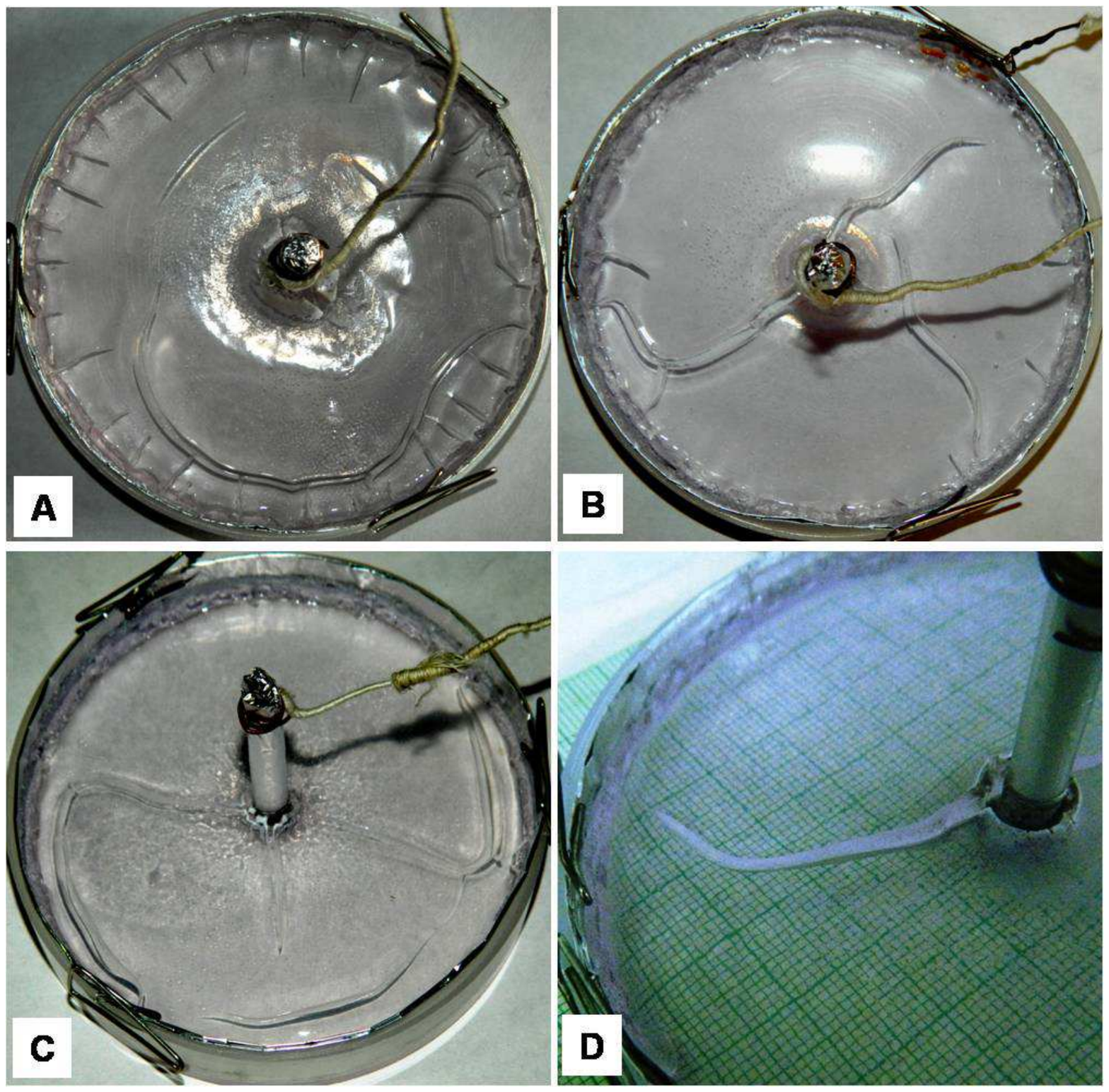}
\end{center}
\caption{Figures A, B, C and D show respectively, the crack patterns of 20 V, 40 V, 60 V and the close up view of a crack with curvature (at 30 V, after 4 hr 52 min). They are not the final patterns but taken when the typical characteristics are best visible, for A, B and C the snapshots are taken after 22 hr 47 min, 23 hr 26 min and 7 hr 11 min respectively. In (A) cracks start to form first from peripheral electrode and some cracks appear from central electrode at a later time. In B and C cracks start to appear from central electrode first and then from outer electrode.}
\label{patterns}
\end{figure}

Figure(\ref{patterns} A-C) shows typical patterns for 20 V, 40 V and 60 V applied to the sample, during drying. For relatively lower voltages (20 V, 25 V), cracks start to appear at the periphery  $\sim 85-90$ minutes after the field is turned on. But interestingly, they do not proceed radially towards the center, rather they curve, forming a nearly circular pattern along the boundary. At a later time some cracks also form from the central electrode.  As the voltage is increased to 30 V or higher, cracks appear initially from the center, in addition to some from the periphery at a later time. Those from the center do not however, run straight in the radial direction, but tend to curve around. The curvature increases with the applied voltage. A close-up of the curved crack is shown in figure(\ref{patterns} D). We could not go higher than 70 V, due to sparking near the central electrode. 

\begin{figure}[h]
\begin{center}
\includegraphics[width=12.0cm, angle=0]{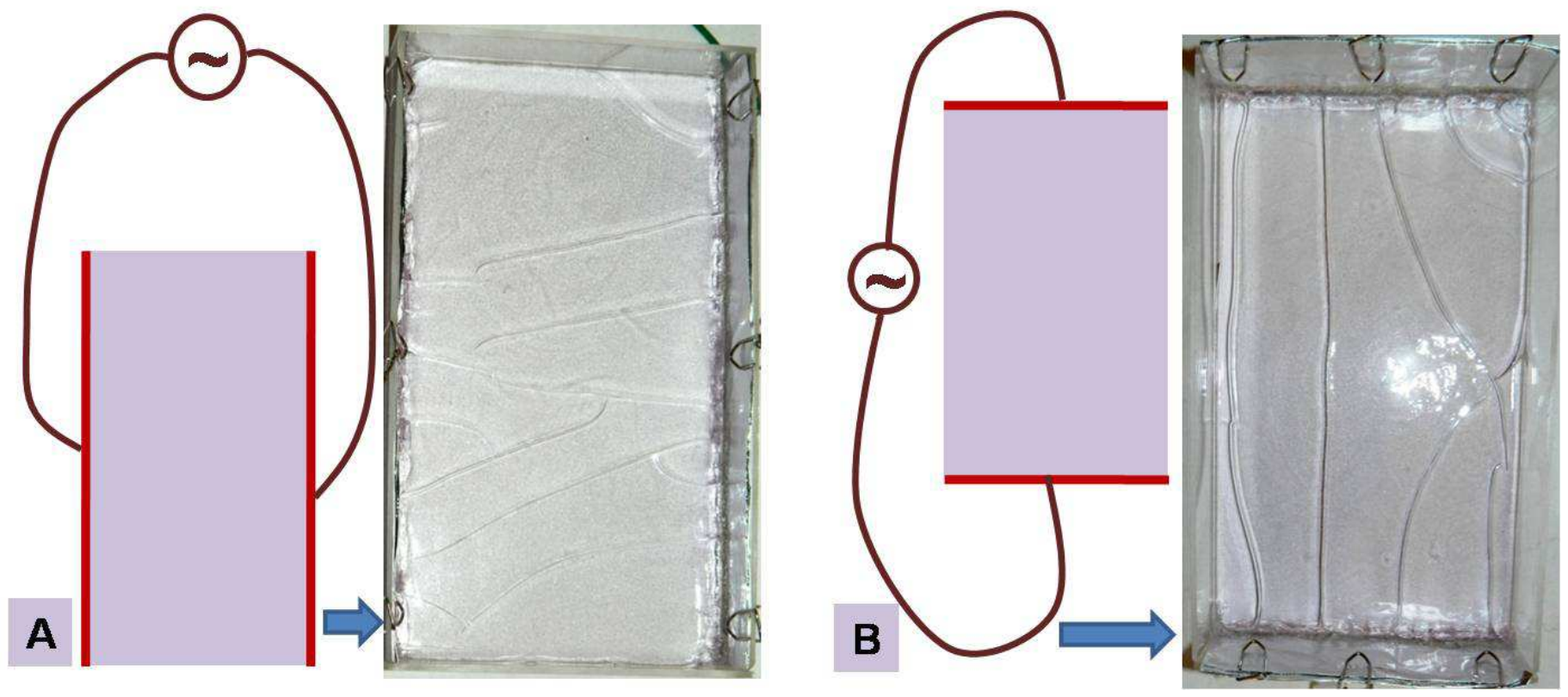}
\end{center}
\caption{Crack patterns for AC field in rectangular geometry of size 20 cm x 11.8 cm. Figure (A) shows the crack patterns, when the field is applied width-wise along with a schematic of the set up. Figure (B) shows the crack patterns, when the field is applied length-wise along with the schematic of the set up.}
\label{rect}
\end{figure}

 If we compare the crack growth behaviour in AC to the DC results\cite{tajkera}, the difference is striking. In DC fields, when the center electrode in positive, cracks start at the center and grow radially outward, whereas with the center negative, growth of concentric cross-radial cracks is favoured, with short cracks growing inward from the outer periphery. For AC field studies in the same cylindrical geometry, it is the {\it strength} of the AC field that is responsible for the different patterns, rather than the polarity of the electrodes. An unequal thickening of the gel was observed on application of DC field with the gel perceptibly thicker at the negative electrode than at the positive electrode. This kind of preferential thickening of the gel was not observed on application of AC. Further in the case of DC, water always seeped out of the gel near the negative electrode and cracks started from the positive electrode where the gel dried faster. In the case of AC field, water was observed to seep out always from the central electrode and the first appearance of cracks was mainly decided by the magnitude of the applied voltage. If AC is applied in a uniform rectangular geometry figure(\ref{rect}), cracks grow preferably along the field direction, irrespective of whether the field is directed along the length or breadth of the rectangular box. So it appears that the inhomogeneity of field distribution produced in the circular arrangement is responsible for the peculiarities in pattern formation.

To quantify the results we measure (1) the number of cracks at the inner and outer electrodes (2) the radius of curvature of the curving  cracks from the centre and (3)the time of first crack appearance, all as functions of the applied voltage and (4) the average speed of crack propagation in some cases. Another aspect of the {\it dynamics} of crack propagation is studied as well. Since the laponite gel is transparent, the crack tip opening and its motion can be photographed quite clearly. We can thus compare the direction of crack motion with the direction of parallel striation marks left on the fracture surface after the cracking process completes.

\begin{figure}[h]
\begin{center}
\includegraphics[width=12.0cm, angle=0]{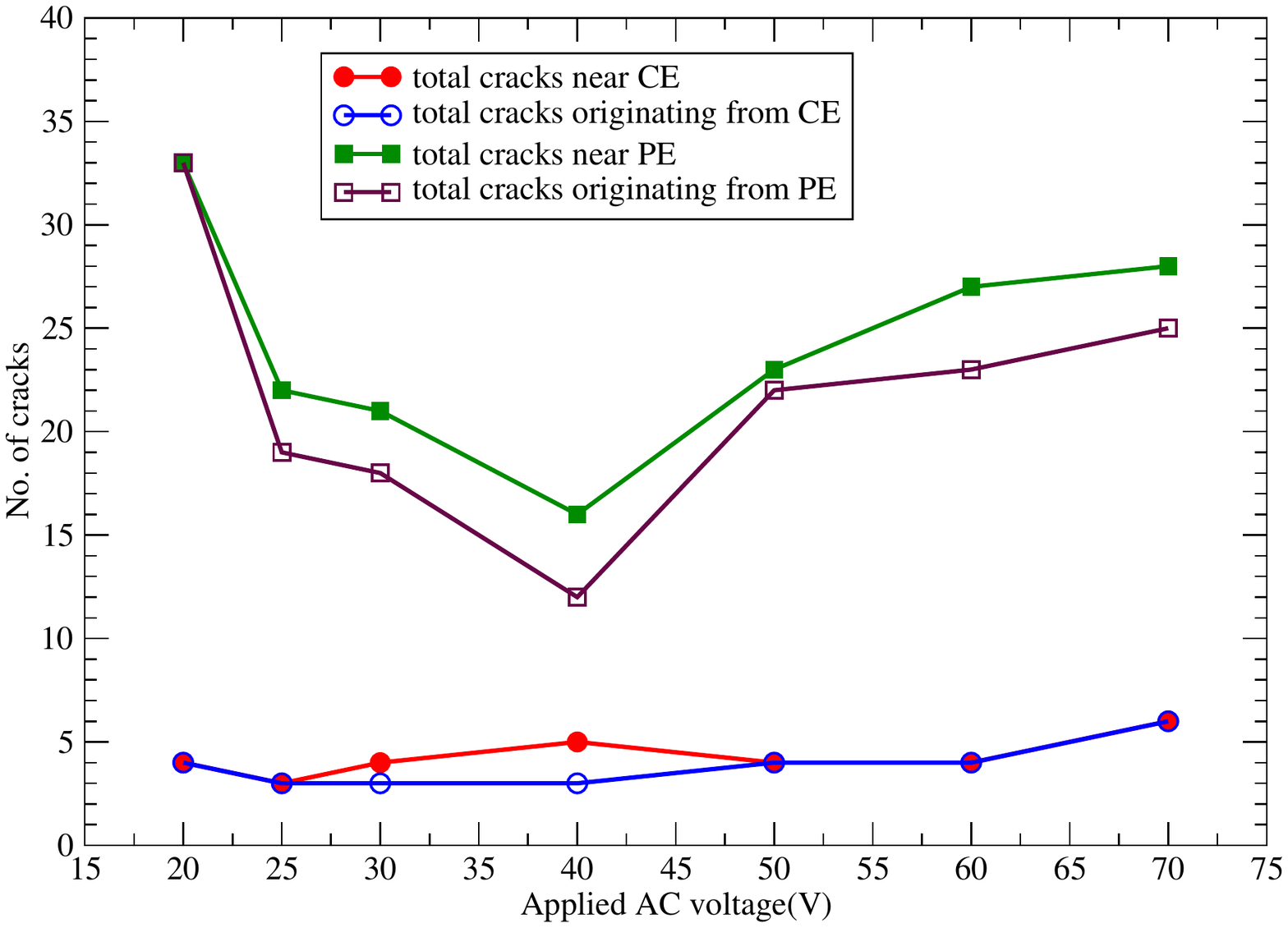}
\end{center}
\caption{Crack number vs applied AC voltage. Red solid circles represent total number of cracks counted at the central electrode (CE) defined as $N_{in}$. Blue open circles represent total number of cracks {\it originated} from the central electrode (CE) defined as $N(O)_{in}$. Green solid squares represent total number of cracks counted at the peripheral electrode (PE) defined as $N_{out}$. Maroon open squares represent total number of cracks originated from peripheral electrode (CE) defined as $N(O)_{out}$.}
\label{cr-num-voltage}
\end{figure}

Figure(\ref{cr-num-voltage}) shows plots of the number of cracks counted at the outer boundary -  $N_{out}$  and the number counted at the central electrode - $N_{in}$ against the voltage. Out of these cracks
some did not originate at the respective electrode, but have arrived there from some other points of the sample, omitting these we count the cracks which {\it originate} at the outer or inner boundary. These are labelled as $N(O)_{out}$ and $N(O)_{in}$ and shown separately on the graph. We see that the number of cracks at the outer electrode first falls with voltage and then increases again with a minimum near 40 V. The number at the inner electrode is much lower $\sim 4 - 6$, compared to $\sim 12 - 32$ at the outer and varies very little with voltage. 

\begin{figure}[h]
\begin{center}

\includegraphics[width=12.0cm, angle=0]{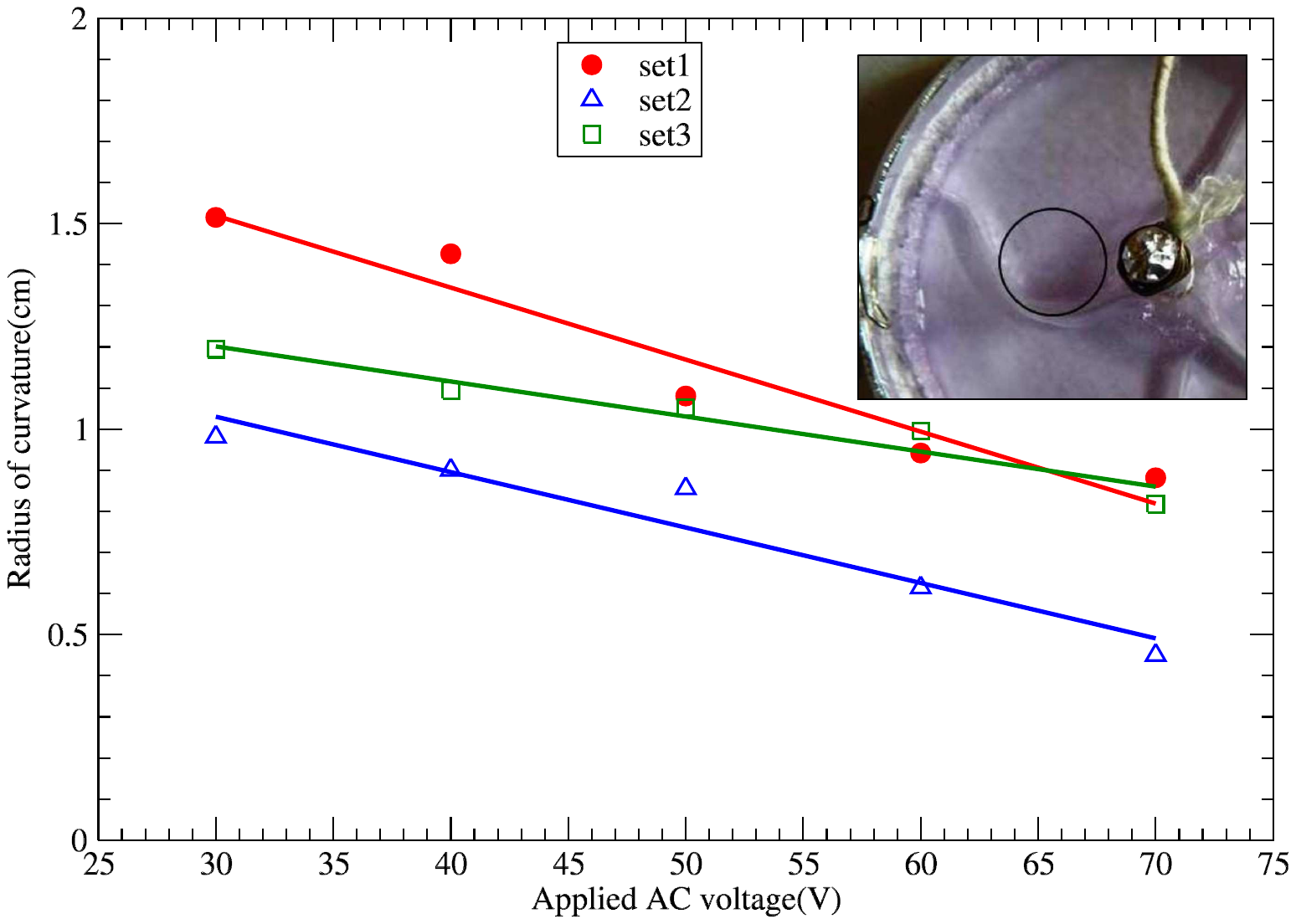}
\end{center}
\caption{Radius of curvature of the crack at the point of strongest curvature vs. applied AC voltages for three different sets. The method of determining the radius of curvature ($R_c$) is shown in the inset. The points in all sets are the experimental data and the solid lines are the best linear fit. The slopes of the fitted straight lines in set1, set2 and set3 are 0.017, 0.013 and 0.009 respectively and are negative in sign. The inset shows how the radius of curvature is measured by fitting a circle.}
\label{radius}
\end{figure}

At higher voltages the cracks starting from the center bend around, the curvature increases with voltage and the distance from the central electrode, where the crack starts to curve decreases. We approximate the radius of curvature ($R_c$) by fitting a circle at the point where the curvature is strongest.  Figure(\ref{radius}) shows how $R_c$ varies with the alternating voltage and the method of determining $R_c$ is shown in the inset. Results for the 3 sets shown are observed under different ambient conditions. For set 1 - temperature is in the range 30 - 32$^{o}C$ and relative humidity is 45 -64 \%, for sets 2 and 3 these are 22 - 26$^{o}C$ , 45 - 60 \% and 26 - 29$^{o}C$, 35 - 44 \% respectively. It seems that in general curvature is less at higher temperature.

At 20 V it takes $\sim 87 min$ for the first crack to appear. As the applied AC voltage increases, the time of first crack appearance decreases rapidly at first, then slower and at 70 V, it becomes $\sim 2 min$. 

\begin{figure}[h]
\begin{center}
\includegraphics[width=12.0cm, angle=0]{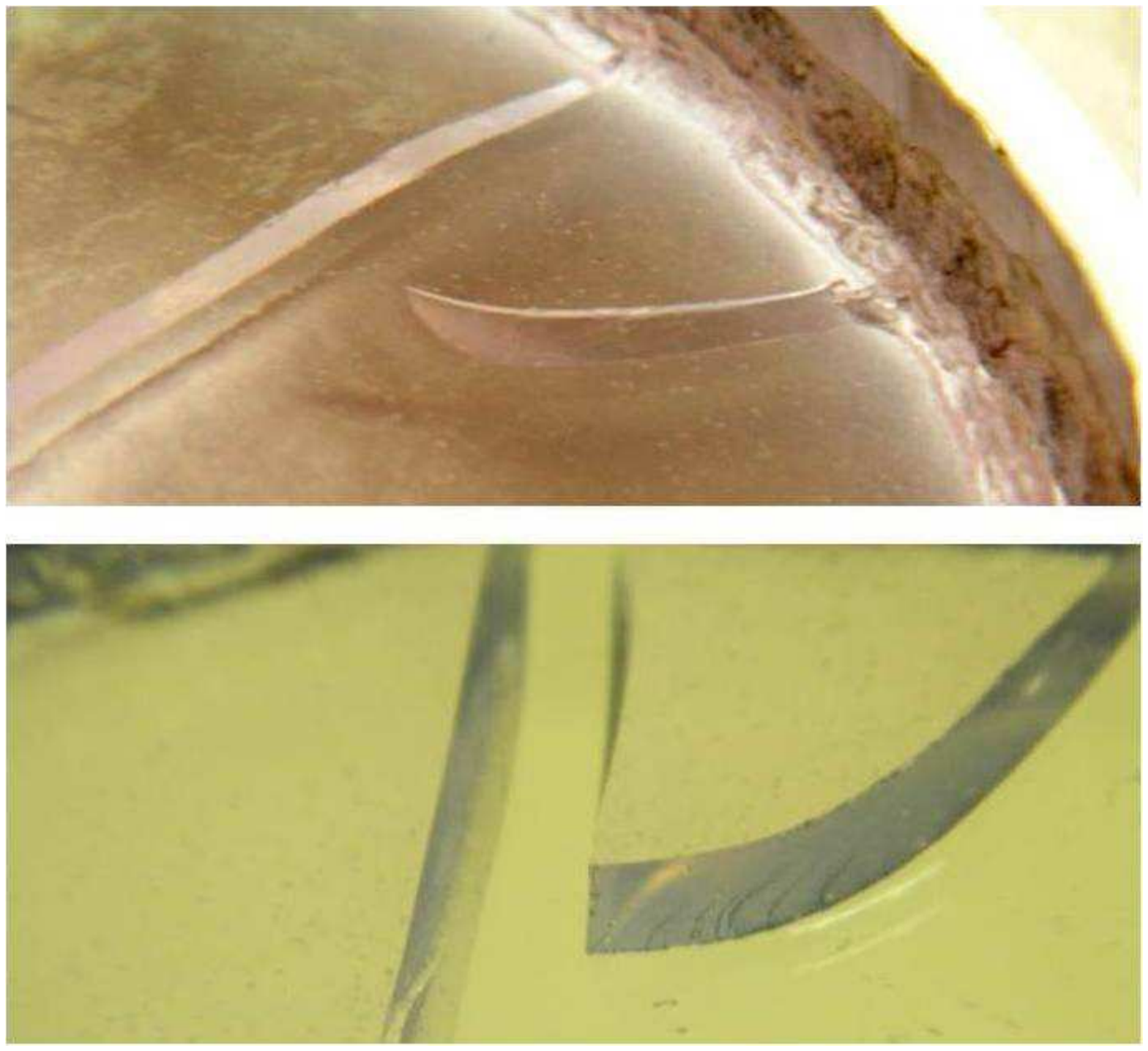}
\end{center}
\caption{Upper figure shows a growing crack tip. This crack clearly started at the upper surface of the laponite layer. Lower Figure shows the striation marks left on the fracture surface by the upper crack after it has grown and met the larger crack seen in the left of the upper figure. The dried gel on the near side of the crack has been removed to show the fracture surface clearly.}
\label{openning}
\end{figure}

A snapshot of a crack tip during growth is shown in the upper part of figure(\ref{openning}). The shape of the front clearly shows that this crack initiates  and  continues to grow along the upper surface of the laponite layer. The lower part of figure(\ref{openning}) shows the fracture surface of the same crack in the upper figure after it has grown completely and intersected the larger crack seen on the left of the upper figure. The gelled material on the other side the crack has been scraped away to provide a clear view of the striations left on the surface. The crack front seen in the upper figure is clearly normal to the directions of the striations as discussed in several texts, e.g. \cite{hull}.

\begin{figure}[h]
\begin{center}
\includegraphics[width=12.0cm, angle=0]{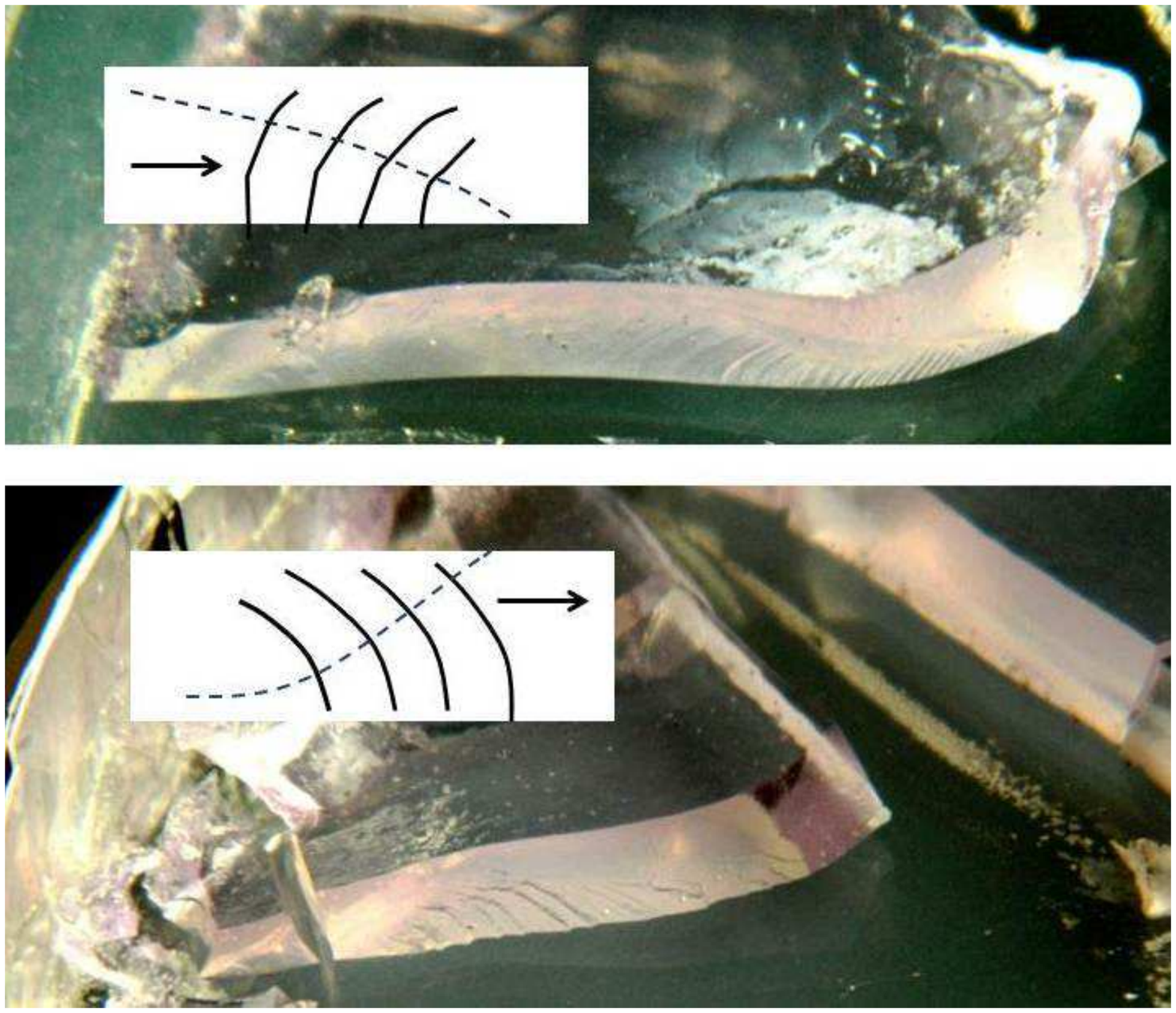}
\end{center}
\caption{Both cracks shown (in upper and lower panels)have grown from left to right, but the stria show opposite orientations. The insets illustrate how we can infer that in the upper figure the crack opens up at lower surface, whereas in the lower figure the crack opens up at upper surface, by drawing the crack front normal to the striations.}
\label{stria}
\end{figure}

 We find opening of cracks along the upper surface to be the usual case, however in some instances a crack may grow from the lower surface in contact with the petri-dish. It is difficult to photograph a crack opening at the lower surface, but the striation marks left on the fracture surface for a crack growing in a specifically observed direction tell us the shape of the crack front and hence the surface where it orginated\cite{hull,so,lucas,aydin1987}. This is demonstrated in the set of figures(\ref{stria}). Both cracks in the upper and lower figures have been observed during growth from left to right. The insets in the upper and lower figures show schematically the directions of the stria and the inferred crack front, the upper(lower) figure striations match a front growing along the lower(upper) surface. In natural environment mud cracks often nucleate at the lower surface of a layer \cite{wien}, though in laboratory experiments they usually open at the upper surface. 
 
The average speed of these cracks can be measured by measuring the displacement of the crack tip at discrete time intervals. We place the sample on a  millimeter graph paper to measure the displacement. Development of several cracks have been recorded and we find that average speeds vary considerably but are found to be within the range 0.5 mm/min (8.333 $\mu$m /s) to 9.25 mm/min (154.167 $\mu$m /s).
 
In homogeneous media a crack normally  follows its original fracture plane unless mixed mode loading tends to deflect it\cite{lawn,prashant}. Simulation of desiccation cracking through a spring network model produces cracks following a circular path in a DC electric field under certain conditions\cite{tajkera}. The AC field creates a more complex situation since it is a dynamic perturbation. Our results show cracks bending normal to the field direction, which may be related to the earlier report by Cao and Evans \cite{cao} of cracks growing normal to the field direction in AC. However we do not observe such bending in the rectangular geometry AC field. So our understanding of the process is far from clear.

To conclude, development of cracks in laponite gel in the presence of AC electric fields shows new and exciting results which need more experiments and analytical study for a fuller understanding.
Tailoring of crack patterns using memory effects\cite{nakahara}, magnetic fields\cite{pauchard} and electric fields both DC\cite{mal,tajkera} and AC may have applications like designing circuits\cite{nature}. So the present study may be of practical importance as well as academic interest. Studies with  variation in frequency of the AC field are planned, which are expected to yield more interesting results. The simulation study\cite{tajkera} for DC fields will extended to incorporate AC fields in an attempt to understand the mechanism behind the pattern formation in different situations.

T.K. thanks CSIR for providing a research grant. Authors are grateful to DST and JSPS for supporting this research through a joint Indo-Japan project. Stimulating discussion and interaction with very helpful suggestions from our Japanese collaborators A.Nakahara, S. Kitsunezaki, C. Urabe, N. Ito, O. Takeshi, T. Hatano and Y. Matsuo, is sincerely acknowledged. We thank Sumit Basu, IIT Kanpur, for helpful suggestions and stimulating discussion. Special acknowledgement is due to Rockwood Additives for gifting the samples of Laponite RD.

\end{document}